\documentclass[aps,prd,reprint,superscriptaddress,nofootinbib]{revtex4-2}

\usepackage{amsmath,amssymb,bm}
\usepackage{microtype}
\usepackage{array}
\usepackage[hidelinks]{hyperref}

\newcommand{\defeq}{\mathrel{:=}}
\newcommand{\D}{\mathcal D}
\newcommand{\B}{\mathcal B}
\newcommand{\Omhat}{\widehat\Omega}
\newcommand{\wh}{\widehat\omega}
\newcommand{\dd}{\mathrm d}
\newcommand{\ri}{\mathrm i}
\newcommand{\e}{\mathrm e}

\begin{document}

\title{An Exact No-Rotation Criterion in Circular Stationary Spacetimes}

\author{Xiao-Yin Pan}
\email{panxiaoyin@nbu.edu.cn}
\affiliation{Department of Physics, Ningbo University, Ningbo 315211, Zhejiang Province, China}

\author{Qing-hai Wang}
\email{qhwang@nus.edu.sg}
\affiliation{Department of Physics, National University of Singapore, Singapore 117551, Singapore}

\date{16 October 2026}

\begin{abstract}
We derive an exact no-rotation criterion for circular, stationary, axisymmetric spacetimes from Einstein's field equations. Its integrated form excludes frame dragging when the weighted matter contribution is nonpositive, the boundary flux vanishes, and the bulk integrals converge. For opposing momentum sources, a finite-region argument excludes frame dragging in a regular region when its boundary values tend uniformly to zero. Time-independent minimally coupled canonical real scalars illustrate the opposing sign, whereas harmonic complex fields can supply angular momentum. In the Barriola--Vilenkin scalar model, the application covers global helical fields with general meridional amplitudes, including possible higher-winding monopole configurations. A persistent vacuum field at an asymptotically conical end locks the phase pattern to the asymptotic frame, excluding horizonless rotation. A nonzero transverse horizon amplitude also synchronizes the horizon with that frame and makes the exterior nonrotating.
\end{abstract}

\maketitle

\section{Introduction}
\label{sec:intro}

Frame dragging connects the angular momentum of a gravitating system to the motion of local inertial frames, with consequences for orbital motion and light propagation. Kerr black holes and rotating stellar models provide familiar examples \cite{Kerr1963,Paschalidis2017}. A proposed stationary rotating geometry must, however, be supported by its matter source and boundary conditions. Angular dependence of the matter fields describes their spatial pattern, but does not by itself imply matter rotation or supply the angular momentum needed to support frame dragging.

These requirements belong to a broader class of restrictions on equilibrium configurations in general relativity. In asymptotically flat vacuum gravity, Lichnerowicz-type no-soliton results exclude nontrivial, globally regular, strictly stationary configurations, whose time-translation Killing field is timelike everywhere \cite{Lichnerowicz1955}. Extensions exclude nontrivial configurations in this class supported by source-free Maxwell fields and minimally coupled massless scalars without a potential, when the fields themselves inherit the time-translation symmetry \cite{Shiromizu2012}. No-hair theorems impose related restrictions in the presence of a horizon. For example, Bekenstein's theorem restricts minimally coupled scalar hair outside asymptotically flat, static, spherically symmetric black holes \cite{Bekenstein1995}. Staticity results ask a complementary question: can a stationary configuration exist while its geometry is forced to be static? Matter may remain present even when frame dragging is excluded. This motivates a direct test of whether the matter source and boundary contributions can sustain stationary rotation.

Our approach builds on established methods for answering this question. Sudarsky and Wald use the Einstein constraints and integral mass identities to prove black-hole staticity results in Einstein--Yang--Mills and Einstein--Maxwell theory, and discuss extensions to other matter \cite{SudarskyWald1992,SudarskyWald1993}. Staticity and circularity results are also established for symmetry-inheriting scalar fields in asymptotically flat black-hole spacetimes \cite{Heusler1993,Heusler1996}. For rotating perfect fluids, Pareja derives an exact dragging equation, maximum-principle bounds, and a positive integral balance \cite{Pareja2004}.

The main result is an exact no-rotation criterion for circular, stationary, axisymmetric spacetimes, without a slow-rotation or weak-field approximation. It expresses the rotational balance through the local matter angular-momentum projection and keeps all boundary contributions, including the horizon term, explicit. The source-and-boundary formulation covers general matter and black-hole boundaries beyond the localized perfect-fluid setting. For momentum opposing frame dragging, its finite-region form excludes rotation from uniformly vanishing boundary values, without infinite-volume rotational integrals. This permits applications to conical monopole geometries, whose total matter energy diverges.

Although the criterion is independent of the matter Lagrangian, scalar fields illustrate how the source and boundary conditions determine whether rotation is excluded. Minimally coupled real fields that are time independent in the asymptotic frame have momentum opposing any assumed frame dragging, irrespective of their potential. Harmonic complex fields instead have a phase frequency that can supply angular momentum, consistently with rotating boson stars and black holes with synchronized scalar hair \cite{Liebling2023,Yoshida1997,Herdeiro2014}. The relevant distinction is the time dependence allowed by stationary stress, rather than the real or complex representation of the fields \cite{Smolic2015,Smolic2017}. For a single harmonic field, the same angular-momentum equation also bounds the local frame-dragging rate.

The Barriola--Vilenkin (BV) global monopole provides a further application. Its triplet of real fields approaches a nonzero angular vacuum profile at infinity, producing an asymptotically conical geometry \cite{BV1989}. Our previous analysis established an obstruction to rotation using a restricted radial scalar deformation and an asymptotic inverse-radius expansion \cite{Lu2026}. The present criterion extends the no-rotation conclusion throughout the connected region, allowing general meridional amplitudes and a common internal rotation that preserves stationary, axisymmetric stress. The persistent field at infinity fixes the phase-pattern rate to the asymptotic frame, thereby excluding horizonless frame dragging. As a subsidiary result, we examine when the stress symmetries require this helical field form. The same no-rotation mechanism extends to higher-winding fields in the BV scalar model with smooth angular vacuum profiles and a conical end whose normal-observer energy density vanishes at infinity. Higher-charge global monopole configurations have also been studied in flat space, including their stability \cite{GudnasonEvslin2015}. Our extension constrains possible self-gravitating solutions without establishing their existence or stability.

Black holes make the role of boundaries particularly clear. A spinning horizon can support frame dragging even when the exterior contains no matter, as in Kerr. For harmonic scalar hair with nonzero horizon amplitude, regularity synchronizes the field's phase evolution with the horizon, while decay at infinity allows both to rotate relative to the asymptotic frame \cite{Herdeiro2014,Smolic2017}. The BV monopole behaves differently: its persistent asymptotic field fixes the phase-pattern rate to that frame. If its transverse amplitude is nonzero on the horizon, synchronization then makes the horizon nonrotating and the criterion excludes frame dragging throughout the exterior. For unit winding, the nonrotating geometry remains compatible with the known static monopole hair \cite{Nucamendi2000,Watabe2004}.

Section~\ref{sec:balance} derives the rotational balance and states the criterion with its boundary conditions. Sections~\ref{sec:real} and \ref{sec:complex} illustrate the contrasting scalar source signs, and Sec.~\ref{sec:BV} applies the criterion to helical monopole and multimonopole fields. Section~\ref{sec:BH} treats horizon contributions and the scalar applications with a black hole. Section~\ref{sec:conclusions} summarizes the conclusions and scope. The appendices give the boundary proofs, scalar-decay estimate, conditional field classification, monopole asymptotics, and horizon estimates in the order of the main-text discussion.

We use signature $(-,+,+,+)$, set $c=\hbar=1$, and retain Newton's constant $G$. 

\section{Geometry and the exact rotational balance}
\label{sec:balance}

\subsection{What is meant by rotation}
We consider stationary, axisymmetric, circular spacetimes.Stationarity gives a time-translation symmetry, and axisymmetry gives a rotational symmetry with closed orbits of period $2\pi$. Their commuting Killing fields are $\bm k\defeq\partial_t$ and $\bm m\defeq\partial_\phi$. Circularity allows coordinates in which the time and azimuthal directions have no metric cross terms with the two meridional directions. The rotational balance below is derived within this class of geometries.

A generic line element of such a spacetime takes the form
\begin{equation}
	\dd s^2 = -N^2\dd t^2 + q_{AB}\dd x^A\dd x^B + \Phi^2(\dd\phi-\Omega\dd t)^2 .
\label{eq:metric}
\end{equation}
The two coordinates $x^A$ describe the meridional geometry, that is, the spatial geometry in a plane through the rotation axis, with positive-definite metric $q_{AB}$. In a common radial and polar coordinate chart, $x^A=(r,\theta)$. All coefficients depend only on $x^A$, with $N>0$ away from a horizon and $\Phi\defeq\sqrt{g_{\phi\phi}}$. The lapse $N$ converts coordinate time to the proper time of observers normal to a constant-$t$ slice. Their four-velocity is
\begin{equation}
	u^\mu\defeq\frac{k^\mu+\Omega m^\mu}{N},
	\qquad u_\mu m^\mu=0 .
\label{eq:normal}
\end{equation}
They have zero axial angular momentum, yet their coordinate angular velocity is $\Omega=-g_{t\phi}/g_{\phi\phi}$. This is the local frame-dragging rate.

A rigidly rotating coordinate system changes this rate. Under $t'=t$, $\phi'=\phi-\Omega_0t$, one has $\Omega'=\Omega-\Omega_0$. Let $\Omega_\infty$ denote the constant asymptotic value of $\Omega$. The frame-dragging rate relative to the physical frame at infinity is
\begin{equation}
	\Omhat\defeq\Omega-\Omega_\infty .
\label{eq:relative}
\end{equation}
Both terms shift by the same constant, so $\Omhat$ is unchanged. Throughout this paper, ``nonrotating'' means $\Omhat=0$, namely the absence of frame dragging relative to the asymptotic frame.

The physical time-translation field is $\bm k_\infty\defeq\bm k+\Omega_\infty\bm m$. When $\Omhat=0$, its metric dual is $k_\infty^\flat=-N^2\dd t$, so it is orthogonal to the constant-$t$ slices and the metric is static. Thus absence of frame dragging establishes staticity within the circular stationary geometries considered here. This statement concerns the geometry and does not require every matter field to be time independent. Zero total angular momentum alone would not establish staticity, since counterrotating contributions can cancel.

The angular-momentum density measured by the normal observers is
\begin{equation}
	j_\phi\defeq-T_{\mu\nu}u^\mu m^\nu
		=-\frac{T_{t\phi}+\Omega T_{\phi\phi}}{N}.
\label{eq:jdef}
\end{equation}
Equivalently, $j_\phi=N T^t{}_\phi$. The mixed component contains the relevant momentum, whereas the covariant component $T_{t\phi}$ alone does not. The axial Killing symmetry also gives a conserved matter current $T^\mu{}_\nu m^\nu$, whose density on the slice is precisely $j_\phi$.

\subsection{Deriving the angular-momentum equation}

In the Arnowitt--Deser--Misner (ADM) formalism \cite{ADM}, the line element is
\begin{equation*}
	\dd s^2=-N^2\dd t^2
		+\gamma_{ij}(\dd x^i+\beta^i\dd t)(\dd x^j+\beta^j\dd t),
\end{equation*}
where $N$ is the lapse, $\gamma_{ij}$ is the induced metric on a constant-$t$ slice, and $\beta^i$ is the shift vector. For the circular stationary metric considered here,
\begin{equation}
	\begin{gathered}
		\gamma_{ij}\dd x^i\dd x^j
		=q_{AB}\dd x^A\dd x^B+\Phi^2\dd\phi^2,\\
		\beta^\phi=-\Omega,\qquad \beta^A=0 .
	\end{gathered}
\label{eq:spatial}
\end{equation}
Let $D_i$ be the covariant derivative of $\gamma_{ij}$, and $\D_A$ that of $q_{AB}$. With the extrinsic-curvature convention
\begin{equation}
	K_{ij}\defeq\frac{D_i\beta_j+D_j\beta_i-\partial_t\gamma_{ij}}{2N},
\label{eq:K}
\end{equation}
the Einstein momentum constraint is \cite{Gourgoulhon2007}
\begin{equation}
	D_j(K^j{}_i-\delta^j_iK)=8\pi G j_i .
\label{eq:constraint}
\end{equation}
Only the mixed meridional--azimuthal components of $K_{ij}$ survive. Since $\beta_\phi=-\Phi^2\Omega$,
\begin{equation}
	\begin{aligned}
		D_A\beta_\phi&=-\Phi^2\partial_A\Omega
		-\Phi\Omega\partial_A\Phi, \\
		D_\phi\beta_A&=\Phi\Omega\partial_A\Phi .
	\end{aligned}
\label{eq:shiftderivatives}
\end{equation}
The derivatives of $\Phi$ cancel, leaving
\begin{equation}
	K_{A\phi}=-\frac{\Phi^2}{2N}\D_A\Omega,\qquad K=0 .
\label{eq:Kphi}
\end{equation}
Using $\sqrt\gamma=\Phi\sqrt q$, the mixed divergence becomes
\begin{equation}
	\begin{aligned}
		D_jK^j{}_\phi
		&=\frac{1}{\Phi\sqrt q}
		\partial_A(\Phi\sqrt q\,K^A{}_\phi)\\
		&=-\frac{1}{2\Phi}\D_A\left(\frac{\Phi^3}{N}\D^A\Omega\right).
	\end{aligned}
\label{eq:divergence}
\end{equation}
Here the two additional mixed-connection terms cancel, as can be checked from $\Gamma^\phi_{A\phi}=\partial_A\ln\Phi$ and $\Gamma^A_{\phi\phi}=-\Phi\D^A\Phi$. The $\phi$ component of Eq.~\eqref{eq:constraint} is consequently
\begin{equation}
	\D_A(W\D^A\Omega)=-16\pi G\,\Phi j_\phi,
	\qquad
	W\defeq\frac{\Phi^3}{N}.
\label{eq:rotation}
\end{equation}
This equation is exact: no expansion in angular velocity, field strength, or distance has been used.\footnote{OpenAI Codex with GPT-5.6 Sol and GPT-6.1 Sol assisted with algebraic checks of the momentum projections, the helical and higher-winding BV extensions, and the near-horizon power counting. The authors set the analytical assumptions and independently verified the resulting expressions against the Einstein and scalar equations.} Away from the axis and a horizon, $W>0$. The equation remains valid where $\bm k_\infty$ becomes spacelike, as long as $N>0$ and the spatial metric is positive definite. Thus an ergoregion does not invalidate the angular-momentum equation.

\subsection{The no-rotation criterion and its boundary conditions}

Since $\Omega_\infty$ is constant, Eq.~\eqref{eq:rotation} also holds for $\Omhat$. Multiplication by $\Omhat$ and integration over a connected meridional region $\B$ give
\begin{align}
	\int_\B W|\D\Omhat|^2\dd V_q
	={}&16\pi G\int_\B\Phi\Omhat j_\phi\dd V_q+B_{\partial\B},
\label{eq:balance}\\
	B_{\partial\B}
	\defeq{}&\oint_{\partial\B}\Omhat W s^A\D_A\Omhat\,\dd S_q .
\label{eq:flux}
\end{align}
Here $\dd V_q\defeq\sqrt q\,\dd x^1\dd x^2$, $\dd S_q$ is the boundary line element, and $s^A$ is the outward unit normal. We first integrate over a finite region bounded away from the axis, center, and any horizon, then take the limits toward these boundaries. The axis and regular-center terms vanish, while any nonzero horizon contribution remains in the balance. All physical boundaries, including additional asymptotic ends, contribute to the boundary term.

The left side is nonnegative. It measures spatially varying frame dragging and vanishes only when $\Omhat$ is constant. From Eq.~\eqref{eq:Kphi} its relation to the standard positive quantity on a spatial slice is
\begin{equation}
	\int_{\Sigma_t}N K_{ij}K^{ij}\dd V_\gamma
	=\pi\int_\B W|\D\Omhat|^2\dd V_q,
\label{eq:Knorm}
\end{equation}
where the slice integral includes the full $2\pi$ azimuthal range. This connects the balance to the positive integral used in staticity arguments \cite{SudarskyWald1993,Heusler1996}. For an asymptotically flat perfect-fluid star with vanishing boundary term, Eq.~\eqref{eq:balance} reduces to the dragging balance derived in Ref.~\cite[Sec.~V]{Pareja2004}. Here the matter contribution is expressed through $j_\phi$, and the boundary term is retained explicitly.

\emph{No-rotation criterion.} Consider a connected region with $N>0$, positive spatial metric, and a fixed asymptotic frame with $\Omhat\to0$. Assume that both bulk integrals in Eq.~\eqref{eq:balance} converge and that the total boundary contribution vanishes in the limit as the integration region approaches the full spacetime region. Then
\begin{equation}
	\int_\B\Phi\Omhat j_\phi\dd V_q\leq0
	\quad\Longrightarrow\quad\Omhat=0.
\label{eq:criterion}
\end{equation}
The balance forces the gradient to vanish. Connectedness and the asymptotic value fix the constant. The geometry is therefore nonrotating relative to that frame. No particular matter Lagrangian enters this result. A nonpositive $\Omhat j_\phi$ at every point guarantees the required integral sign, but positive and negative regions may also give a nonpositive total. The condition involves both the source and the geometry. It is not an energy condition on matter alone. Its practical use is strongest when the matter equations or field ansatz fix the sign for any assumed $\Omhat$, before solving for the geometry. Section~\ref{sec:real} provides such a source. For other matter, the criterion can instead serve as a necessary consistency check on a proposed solution.

A regular center and axis contribute no flux, whereas a horizon or an imposed outer boundary can contribute to the rotational balance. These contributions must be evaluated before applying Eq.~\eqref{eq:criterion}. More generally, rotation is excluded whenever the combined matter and boundary contribution on the right of Eq.~\eqref{eq:balance} is nonpositive. Vacuum Kerr illustrates the role of boundaries: the horizon supplies the positive balance even though $j_\phi=0$ outside the hole. Section~\ref{sec:BH} evaluates the horizon contribution. Appendix~\ref{app:boundaries} gives the axis and center estimates and the decay rates that make the outer flux vanish.

With a vanishing boundary term, nontrivial frame dragging therefore requires a positive weighted matter contribution. This means that the matter angular-momentum density must have the same sign as $\Omhat$ somewhere in the region. This is a necessary condition for rotation, not a guarantee that a rotating solution exists: the remaining Einstein equations and the matter equations must also be satisfied.

\subsection{A boundary-value argument without integrals at infinity}
\label{sec:finite}

A matter momentum that opposes frame dragging at every point allows a no-rotation argument without integrals over the entire unbounded region. Write $z\defeq\Omhat$. For the boundary-value arguments, we consider a connected circular region with $N>0$, a positive spatial metric, and smooth fields and metric in regular coordinates. We assume that $z$ is bounded, the axis is regular, and any center is regular. Suppose that
\begin{equation}
	z j_\phi\leq0
\label{eq:pointwisesign}
\end{equation}
throughout the region. Consider a connected finite meridional region, bounded at large distance and, for a black-hole exterior, just outside the horizon. Let $\delta$ be the largest $|z|$ on all imposed outer and inner boundaries. Equation~\eqref{eq:rotation} then gives
\begin{equation}
	|z|\leq\delta\quad\hbox{throughout the finite region}.
\label{eq:bound}
\end{equation}
Thus frame dragging cannot exceed its largest boundary value in magnitude. Appendix~\ref{app:comparison} proves this bound and evaluates the axis and center limits.

For the horizonless applications below, we assume no additional physical boundary and $\Omhat\to0$ uniformly at infinity. These boundary conditions make $\delta\to0$ as the finite region grows, giving $\Omhat=0$ throughout the connected region. If other physical boundaries are present, the same no-rotation conclusion requires $\Omhat$ to tend uniformly to zero there as well. The argument takes the limit of a boundary-value bound, so it requires neither a specified decay rate for $\Omhat$ nor convergence of infinite-volume rotational integrals.

A useful source form that guarantees Eq.~\eqref{eq:pointwisesign} is
\begin{equation}
	j_\phi=-\frac{\Omhat}{N}S,
	\qquad 
	S\geq0 .
\label{eq:signsource}
\end{equation}
Equation~\eqref{eq:rotation} then becomes
\begin{equation}
	\D_A(W\D^Az)=cz,
	\qquad
	c\defeq16\pi G\frac{\Phi S}{N}\geq0 .
\label{eq:comparisonPDE}
\end{equation}
The time-independent scalar examples below have this structure. Vacuum regions are included by taking $S=0$.

\subsection{Matter rotation and frame dragging}

The angular velocity of matter need not coincide with the frame-dragging rate. For a circular perfect fluid, $U^\mu=U^t(k^\mu+\Omega_{\rm f}m^\mu)$, where $\Omega_{\rm f}$ is the fluid's coordinate angular velocity. The normal observers rotate at rate $\Omega$ and measure
\begin{equation*}
	\begin{aligned}
		v&\defeq\frac{\Phi}{N}(\Omega_{\rm f}-\Omega),
		\\
		\Gamma&\defeq-u_\mu U^\mu=NU^t=\frac{1}{\sqrt{1-v^2}}.
	\end{aligned}
\end{equation*}
Here $v$ is the fluid's azimuthal velocity measured by the normal observers, and $\Gamma$ is the corresponding Lorentz factor.

For the perfect-fluid stress tensor $T_{\mu\nu}=(\rho_{\rm f}+p_{\rm f})U_\mu U_\nu +p_{\rm f}g_{\mu\nu}$, the projection \eqref{eq:jdef} gives
\begin{equation}
	j_\phi
	= (\rho_{\rm f} + p_{\rm f}) \Gamma^2 \frac{\Phi^2}{N} (\Omega_{\rm f}- \Omega).
\label{eq:fluidj}
\end{equation}
The metric-proportional pressure term drops out because $u_\mu m^\mu=0$. For $\rho_{\rm f}+p_{\rm f}\geq0$, the momentum sign therefore follows the fluid's motion relative to the normal observers. A fluid rotating faster than these observers can supply a positive contribution to the rotational balance when $\Omhat>0$.

If instead the fluid is at rest in the asymptotic frame, $\Omega_{\rm f}=\Omega_\infty$, then
\begin{equation*}
	\Omhat j_\phi 
	= -\frac{(\rho_{\rm f} + p_{\rm f}) \Gamma^2\Phi^2}{N} \Omhat^2 \leq0.
\end{equation*}
For the regular horizonless configurations described in Sec.~\ref{sec:finite}, the finite-region argument forces $\Omhat=0$. The opposing sign rules out a stationary rotating equilibrium. It does not describe friction or dynamical spin-down.

\section{Scalar examples: time-independent real fields}
\label{sec:real}

\subsection{Why an angular pattern does not supply rotation}

Consider any number of minimally coupled real fields with
\begin{equation}
	\mathcal L_{\rm s}
	=-\frac12\nabla_\mu\bm\varphi\cdot\nabla^\mu\bm\varphi - V(\bm\varphi),
\label{eq:realL}
\end{equation}
where the dot denotes the Euclidean inner product over scalar components, $\bm a\cdot\bm b\defeq\sum_I a^I b^I$, with repeated internal indices summed. The kinetic term is canonical and there is no explicit curvature coupling in the matter action. Their stress tensor is
\begin{equation}
	T_{\mu\nu}
	=\partial_\mu\bm\varphi\cdot\partial_\nu\bm\varphi
	+g_{\mu\nu}\mathcal L_{\rm s}.
 \label{eq:realT}
\end{equation}
The metric-proportional term contributes nothing to $j_\phi$ because $u^\mu m_\mu=0$. Direct substitution gives
\begin{equation}
	j_\phi
	=-\frac1N (\partial_t\bm\varphi + \Omega\partial_\phi\bm\varphi) \cdot \partial_\phi \bm\varphi.
 \label{eq:realjgeneral}
\end{equation}

Assume the fields are time independent with respect to the physical time-translation field $\bm k_\infty$ throughout the region:
\begin{equation}
	(\partial_t+\Omega_\infty\partial_\phi)\bm\varphi=0 .
\label{eq:inherit}
\end{equation}
This is a condition on the matter, not a condition that can always be imposed by choosing coordinates. It yields
\begin{equation}
	j_\phi = -\frac{\Omhat}{N}|\partial_\phi\bm\varphi|^2,
	\qquad
	\Omhat j_\phi=-\frac{\Omhat^2}{N} |\partial_\phi\bm\varphi|^2\leq0 .
\label{eq:realj}
\end{equation}
The scalar components may still depend on $\phi$. If $\Omhat\ne0$, the normal observers move through this angular pattern and measure 
$u^\mu\partial_\mu\bm\varphi = (\Omhat/N)\partial_\phi\bm\varphi$. The resulting angular-momentum density opposes frame dragging, rather than supplying the positive contribution needed to support it when the boundary term vanishes. The potential affects the field profiles and geometry, but drops out of this momentum projection. The source sign therefore holds for any $V$ in Eq.~\eqref{eq:realL}.

With $S=|\partial_\phi\bm\varphi|^2$, the source has the form of Eq.~\eqref{eq:signsource}. Therefore, the finite-region argument gives $\Omhat=0$ throughout the region. Thus a time-independent angular pattern cannot support stationary frame dragging under these boundary conditions.

\subsection{The BV monopole as an example}

For the BV triplet the potential is
\begin{equation}
	V(\bm\chi)=\frac{\lambda}{4}(|\bm\chi|^2-\eta^2)^2,
	\qquad \lambda>0,\quad\eta>0 .
\label{eq:BVpotential}
\end{equation}
The vacuum states form a sphere $|\bm\chi|=\eta$. The familiar hedgehog configuration is $\bm\chi=\eta h(r)\hat{\bm r}$, with $h\to1$ at infinity \cite{BV1989}. Its internal direction follows the spatial angular coordinates, but the field is time independent. Working in coordinates nonrotating at infinity, $\Omega_\infty=0$, consider the broader time-independent monopole profile
\begin{equation}
	\bm\chi=(F(r,\theta)\cos\phi,\ F(r,\theta)\sin\phi,\ H(r,\theta)).
\label{eq:staticBV}
\end{equation}
Its stress tensor is axisymmetric even though the first two scalar components depend on $\phi$. The usual hedgehog is recovered with $F=\eta h(r)\sin\theta$ and $H=\eta h(r)\cos\theta$. The restricted radial deformation used in Ref.~\cite{Lu2026} instead takes $F=\eta h(r)\zeta(r)\sin\theta$, with $\zeta(r)$ a radial deformation function, while retaining the same $H$.

For arbitrary meridional amplitudes $F$ and $H$, $|\partial_\phi\bm\chi|^2=F^2$, so
\begin{equation}
	j_\phi=-\frac{\Omega F^2}{N}.
\label{eq:staticBVj}
\end{equation}
Thus $\Omhat j_\phi=-\Omega^2F^2/N\leq0$, independently of the radial or angular dependence of the amplitudes.

For the usual monopole asymptotics, angular gradients produce stress of order $\eta^2/r^2$ and an asymptotically conical geometry with deficit parameter $\Delta\defeq8\pi G\eta^2<1$ \cite{BV1989}. The total matter energy therefore diverges, but the finite-region argument still excludes frame dragging.

\section{Scalar examples: harmonic complex fields}
\label{sec:complex}

\subsection{A stationary source with a moving phase}

Stationary stress need not arise from time-independent fields. An internal phase rotation can leave the stress unchanged \cite{Smolic2015}. To illustrate this mechanism, consider
\begin{equation}
	\Psi\defeq\frac{\varphi^1+\ri\varphi^2}{\sqrt2},\qquad
	\mathcal L_{\rm s}=-\nabla_\mu\Psi^*\nabla^\mu\Psi-U(s).
\label{eq:complexL}
\end{equation}
Here $s\defeq|\Psi|^2$. Variation gives
\begin{align}
	T_{\mu\nu}&=\partial_\mu\Psi^*\partial_\nu\Psi + \partial_\nu\Psi^*\partial_\mu\Psi  + g_{\mu\nu}\mathcal L_{\rm s},\label{eq:complexT}\\
	\Box\Psi&=U'(s)\Psi,\qquad U'(s)\defeq\frac{\dd U}{\dd s}.
\label{eq:complexEOM}
\end{align}
Take a harmonic mode
\begin{equation}
	\Psi=f(x^A)\e^{\ri\Theta},\qquad
	\Theta\defeq n\phi-\omega t,\qquad n\in\mathbb Z ,
\label{eq:complexansatz}
\end{equation}
with real $f$ and constant $\omega$. Single-valuedness under $\phi\to\phi+2\pi$ fixes the integer winding $n$. The phase cancels in the stress because
\begin{equation}
	2\operatorname{Re}(\partial_\mu\Psi^*\partial_\nu\Psi)
	=2\partial_\mu f\,\partial_\nu f + 2f^2\partial_\mu\Theta\,\partial_\nu\Theta .
\label{eq:phasecancel}
\end{equation}
Thus the stress tensor is stationary and axisymmetric. Moreover, $T_{At}=T_{A\phi}=0$, consistently with the assumed circular metric. For $n\ne0$, surfaces of constant phase advance at angular rate $\omega/n$. This phase-pattern rate alone does not determine the field's angular-momentum density, which must be calculated from the scalar stress tensor, as derived below.

Under the rigid coordinate change $t'=t$, $\phi'=\phi-\Omega_0t$, the arbitrary constant $\Omega_0$ gives $\omega'=\omega-n\Omega_0$ and $\Omega_\infty'=\Omega_\infty-\Omega_0$. Thus the frequency relative to the asymptotic frame is unchanged:
\begin{equation}
	\wh\defeq\omega-n\Omega_\infty,\qquad
	\omega-n\Omega=\wh-n\Omhat .
\label{eq:frequency}
\end{equation}
Choosing $\Omega_0=\Omega_\infty$ gives coordinates nonrotating at infinity, with $\Omega_\infty'=0$ and $\omega'=\wh$. For the asymptotically flat scalar examples below, let $N_\infty\defeq\lim_{r\to\infty}N>0$. The proper-time frequency at infinity is $\wh/N_\infty$. From $u^\mu\partial_\mu\Psi=-\ri(\omega-n\Omega)\Psi/N$ and $\partial_\phi\Psi=\ri n\Psi$, Eq.~\eqref{eq:jdef} gives
\begin{equation}
	\begin{aligned}
		j_\phi
		&=-2\operatorname{Re}\left[
		(u^\mu\partial_\mu\Psi^*)\partial_\phi\Psi\right] \\
		&=\frac{2n}{N}(\omega-n\Omega)f^2
		=\frac{2n}{N}(\wh-n\Omhat)f^2 .
	\end{aligned}
\label{eq:complexj}
\end{equation}
The factor two follows from the normalization in Eq.~\eqref{eq:complexL}.

The phase frequency supplies an additional term proportional to $\wh$ in the momentum source, so the source need not oppose frame dragging. Substitution into Eq.~\eqref{eq:rotation} gives
\begin{align}
	-\D_A(W\D^A\Omhat)+C\Omhat
	&=32\pi G\,\frac{n\Phi f^2}{N}\wh,
\label{eq:complexdriven}
\end{align}
with $C\defeq32\pi G{n^2\Phi f^2}/{N}\geq0.$ For example, in coordinates with $\Omega_\infty=0$,
$n,\omega>0$ and $0<\Omega<\omega/n$ give $\Omega j_\phi>0$ wherever $f\ne0$. The scalar can therefore supply a positive contribution to the rotational balance. When $\wh=0$, the opposing source sign is recovered. When $n=0$, the angular-momentum density vanishes regardless of the phase frequency. The freedom to support rotation comes from time-dependent fields with stationary stress.

The restriction found for the slow-rotation construction of Ref.~\cite{Kobayashi1994} does not exclude other rotating scalar configurations, including the harmonic modes used in rotating boson stars \cite{Yoshida1997} and the approximate constructions of Ref.~\cite{Delgado2025}.

\subsection{A bound on frame dragging}

For the horizonless boundary conditions of Sec.~\ref{sec:finite}, the same comparison argument bounds the frame dragging supported by the harmonic scalar. For $n\ne0$, define the phase-pattern rate relative to infinity as $\widehat\Omega_{\rm p}\defeq\wh/n$. Equation~\eqref{eq:complexdriven} then reads
\begin{equation}
	-\D_A(W\D^A\Omhat)+C\Omhat=C\widehat\Omega_{\rm p}.
\label{eq:patternPDE}
\end{equation}
Since $C\geq0$, comparison with $0$ and $\widehat\Omega_{\rm p}$ gives
\begin{equation}
	\min(0,\widehat\Omega_{\rm p})\leq\Omhat
	\leq\max(0,\widehat\Omega_{\rm p}).
\label{eq:patternbound}
\end{equation}
Thus frame dragging relative to infinity lies between zero and the field's relative phase-pattern rate. Where frame dragging and the scalar angular-momentum density are both nonzero, Eq.~\eqref{eq:complexj} shows that they have the same sign. The bound constrains any such solution, while its existence and spatial profiles depend on the remaining Einstein and scalar equations. Appendix~\ref{app:comparison} gives the finite-region proof.

\subsection{Phase charge and angular momentum}
\label{sec:phasecharge}

To connect the local momentum source to the total angular momentum, we now relate it to the scalar's conserved phase charge. The same relation will separate the scalar and horizon contributions in Sec.~\ref{sec:BH}.

The global phase symmetry has conserved current
\begin{equation}
	\mathcal J^\mu\defeq-\ri(\Psi^*\nabla^\mu\Psi-\Psi\nabla^\mu\Psi^*),
	\qquad \nabla_\mu\mathcal J^\mu=0 .
\label{eq:current}
\end{equation}
For the harmonic field $\mathcal J^\mu=2f^2\partial^\mu\Theta$. The normal observers measure the charge density
\begin{equation}
	\rho_{\mathcal Q}\defeq-u_\mu\mathcal J^\mu
	=\frac{2(\omega-n\Omega)f^2}{N},\qquad
	j_\phi=n\rho_{\mathcal Q}.
\label{eq:chargedensity}
\end{equation}
For the real-amplitude ansatz, the meridional current vanishes, $\mathcal J^A=0$. A regular localized star therefore has no charge flux through its outer boundary, and its integrated phase charge is conserved. With the slice measure $\dd V_\gamma=\Phi\dd V_q\dd\phi$, we obtain
\begin{equation}
	\mathcal Q\defeq\int_{\Sigma_t}\rho_{\mathcal Q}\dd V_\gamma,\qquad
	J_{\rm s}\defeq\int_{\Sigma_t}j_\phi\dd V_\gamma=n\mathcal Q .
\label{eq:Js}
\end{equation}

To relate $J_{\rm s}$ to the total angular momentum, define $P^j{}_i\defeq K^j{}_i-\delta^j_iK$. Contracting the momentum constraint with the spatial axial Killing field gives
\begin{equation}
	D_j(P^j{}_im^i)=8\pi G j_\phi .
\label{eq:Jdiv}
\end{equation}
The term $P^{ji}D_jm_i$ vanishes because it contracts a symmetric and an antisymmetric tensor. For an asymptotically flat end, the ADM angular momentum is
\begin{equation}
	J\defeq\frac{1}{8\pi G}\lim_{S_R\to\infty}
	\int_{S_R}P^j{}_i m^i r_j\dd A,
\label{eq:Jsurface}
\end{equation}
with outward normal $r_j$ and surface area element $\dd A$. Integrating Eq.~\eqref{eq:Jdiv} over the spatial slice identifies the ADM angular momentum with the scalar contribution:
\begin{equation}
	J=J_{\rm s}=n\mathcal Q .
\label{eq:Jstar}
\end{equation}
The charge $\mathcal Q$ is continuous, so integer winding does not quantize the classical angular momentum. Rotating boson stars realize this relation \cite{Yoshida1997,Liebling2023}.

\subsection{Why localization permits a nonzero frequency}

For a massive field with $U(s)=\mu^2s+O(s^2)$ near $s=0$, where $\mu>0$, exponential decay in an asymptotically flat geometry requires the proper-time frequency at infinity to lie below the scalar mass:
\begin{equation}
	|\wh|<\mu N_\infty .
\label{eq:frequencybound}
\end{equation}
Appendix~\ref{app:decay} derives this condition from the scalar's asymptotic amplitude equation. The mass term then permits a decaying tail even when $\wh\ne0$. Localization therefore restricts the frequency without fixing it to zero.

For $n\ne0$, this allows a phase pattern that rotates relative to infinity and can supply angular momentum through Eq.~\eqref{eq:complexj}. The allowed frequencies and field profiles are determined by the remaining Einstein and scalar equations. The persistent monopole field considered next imposes a stronger restriction on the internal frequency.

\section{A further application: helical monopoles and multimonopoles}
\label{sec:BV}

\subsection{A common internal rotation frequency}

The time-independent monopole of Sec.~\ref{sec:real} already illustrates the criterion. We now allow its first two components to rotate into each other and include arbitrary nonzero integer azimuthal winding:
\begin{equation}
	\Xi\defeq\chi^1+\ri\chi^2   
	=F(x^A)\e^{\ri(n\phi-\omega t)},\qquad 
	\chi^3=H(x^A)
\label{eq:helical}
\end{equation}
with $n\in\mathbb Z\setminus\{0\}$. Here $F,H$ are real, and the constant $\omega$ is the internal time frequency throughout the configuration. The phase-pattern angular rate is $\omega/n$. For this field form, the kinetic inner products and potential are independent of $t$ and $\phi$, so the stress tensor is stationary and axisymmetric. The fields are constant along $\partial_t+(\omega/n)\partial_\phi$, which explains the term ``helical.'' The standard BV hedgehog has $n=1$, while higher winding requires different smooth angular boundary profiles.

We prescribe Eq.~\eqref{eq:helical} globally for the application below. Appendix~\ref{app:symmetry} examines when this form can instead be derived from the stress symmetries, circularity, and a specified monopole boundary profile.

Since $|\partial_\phi\bm\chi|^2=n^2F^2$, the momentum source is
\begin{equation}
	\begin{aligned}
		u^\mu\partial_\mu\bm\chi
		&=\frac{\Omega-\omega/n}{N}\partial_\phi\bm\chi, \qquad
		j_\phi=\frac{n(\omega-n\Omega)}{N}F^2.
	\end{aligned}
\label{eq:BVj}
\end{equation}
The normalized complex representation of the pair is $\Xi/\sqrt2$, so $F^2$ corresponds to $2f^2$ in Eq.~\eqref{eq:complexj}. The term proportional to $\omega$ can supply angular momentum. The monopole boundary condition, examined next, determines whether this frequency freedom remains.

\subsection{Why the monopole boundary fixes the frequency}

We take a persistent angular vacuum profile at infinity,
\begin{equation}
	\begin{gathered}
		F\to F_\infty(\theta),\qquad H\to H_\infty(\theta),\\
		F_\infty^2+H_\infty^2=\eta^2,\\
		N\to N_\infty>0,\qquad \Omega\to\Omega_\infty.
	\end{gathered}
\label{eq:BVasymptotics}
\end{equation}
For monopole charge $n$, choose $H_\infty(0)=\eta$ and $H_\infty(\pi)=-\eta$, with the full triplet smooth at both axis endpoints. Appendix~\ref{app:windingprofiles} gives the charge calculation and examples of smooth higher-winding vacuum profiles. The standard BV profile is recovered for $n=1$, $F_\infty=\eta\sin\theta$, and $H_\infty=\eta\cos\theta$.

We use an asymptotically conical end, allowing angular dependence in the leading metric coefficients. The asymptotic azimuthal shift obeys $\Phi\Omhat\to0$ uniformly, so the physical time-translation field $\bm k_\infty$ is timelike at infinity. The spatial scalar curvature and $K_{ij}K^{ij}$ tend to zero, as shown by the metric falloffs in Appendix~\ref{app:monopole}. The ADM Hamiltonian constraint, with $K=0$, gives
\begin{equation*}
	16\pi G\rho={}^{(3)}R-K_{ij}K^{ij},
\end{equation*}
where ${}^{(3)}R$ is the intrinsic scalar curvature of the spatial metric on a constant-$t$ slice. Both geometric terms vanish at the asymptotic end, so the normal-observer energy density tends to zero. For the canonical BV triplet,
\begin{equation}
	\begin{aligned}
		\rho\defeq{}& T_{\mu\nu}u^\mu u^\nu\\
		={}&\frac{(\omega-n\Omega)^2F^2}{2N^2}
		+\frac12q^{AB}\partial_AF\partial_BF\\
		&\quad+\frac12q^{AB}\partial_AH\partial_BH
		+\frac{n^2F^2}{2\Phi^2}+V(F,H).
	\end{aligned}
\label{eq:BVenergy}
\end{equation}
Here $V(F,H)\defeq\lambda(F^2+H^2-\eta^2)^2/4$ is the BV potential in Eq.~\eqref{eq:BVpotential}, expressed in terms of the meridional amplitudes. The canonical kinetic terms and nonnegative BV potential make every contribution nonnegative. In particular,
\begin{equation}
	\rho\geq\frac{(\omega-n\Omega)^2F^2}{2N^2}.
\label{eq:BVenergybound}
\end{equation}
A frequency $\omega\ne n\Omega_\infty$ would retain a nonzero energy density wherever $F_\infty\ne0$, contradicting the asymptotic geometry. Hence
\begin{equation}
	\omega=n\Omega_\infty.
\label{eq:syncinfty}
\end{equation}
Equivalently, the invariant frequency defined in Eq.~\eqref{eq:frequency} vanishes, $\wh=0$. Hence $k_\infty^\mu\partial_\mu\bm\chi=0$: the boundary locks the phase pattern to the asymptotic frame. This conclusion requires a nonzero transverse amplitude at only one asymptotic angle and no falloff estimates for scalar second derivatives.

The physical distinction from a localized scalar is its persistent asymptotic amplitude. A localized field can make its time-derivative energy vanish by decaying to zero. For the monopole, a nonzero relative frequency would instead leave a nonzero energy density arbitrarily far away, so the asymptotic geometry forces that frequency to vanish.

\subsection{Applying the no-rotation criterion}

Substitution of Eq.~\eqref{eq:syncinfty} into Eq.~\eqref{eq:BVj} gives
\begin{equation}
	\begin{aligned}
		j_\phi&=-\frac{n^2F^2}{N}\Omhat,\\
		\D_A(W\D^A\Omhat)
		&=16\pi G\frac{n^2\Phi F^2}{N}\Omhat.
	\end{aligned}
\label{eq:BVpositivePDE}
\end{equation}
The source has the form of Eq.~\eqref{eq:signsource}, with $S=n^2F^2\geq0$. Then the finite-region argument gives $\Omhat=0$, without requiring finite total monopole energy or convergent infinite-volume rotational integrals. Higher winding introduces the positive factor $n^2$ and leaves the opposing momentum sign unchanged.

For $n=1$, this extends Ref.~\cite{Lu2026} to general meridional amplitudes and excludes frame dragging throughout the connected region, without an inverse-radius expansion. For $|n|>1$, the same conclusion applies to global helical multimonopole solutions. The remaining Einstein and scalar equations determine which solutions exist.

Dynamical stability is a separate question. Gudnason and Evslin found repulsion and instability of charge-two global monopoles in the flat-space canonical model, and stabilized a configuration through a coupling to a Skyrme sector \cite{GudnasonEvslin2015}. 

\section{Black-hole boundaries and the scalar examples}
\label{sec:BH}

\subsection{The horizon contribution to the rotational balance}

A horizon changes the problem because it is an inner boundary that can carry angular momentum. We take a connected regular Killing horizon with generator
\begin{equation}
	\ell^\mu\defeq k^\mu+\Omega_Hm^\mu .
\label{eq:horizongenerator}
\end{equation}
Its angular velocity relative to infinity is $\Omhat_H\defeq\Omega_H-\Omega_\infty$. A nonrotating horizon has $\Omhat_H=0$.

Let $S_\varepsilon$ be a surface just outside the horizon, with spatial unit normal $r^i$ pointing from the hole into the exterior. The outward normal of the exterior integration region is therefore $s^i=-r^i$. Let $\mathcal C_\varepsilon$ be its meridional curve, with $\dd A=\Phi\dd S_q\dd\phi$. Since $K=0$, Eq.~\eqref{eq:Kphi} gives the horizon angular momentum as
\begin{equation}
	\begin{aligned}
		J_H
		&=\frac1{8\pi G}\lim_{\varepsilon\to0}\int_{S_\varepsilon} K_{A\phi}r^A\dd A\\
		&=-\frac1{8G}\lim_{\varepsilon\to0}\int_{\mathcal C_\varepsilon}
		W r^A\D_A\Omega\,\dd S_q .
	\end{aligned}
\label{eq:JHflux}
\end{equation}
Assume that $J_H$ has a finite limit, that $\Omhat\to\Omhat_H$ uniformly, and that the integral of the absolute rotational flux density remains bounded. These limits allow $\Omhat_H$ to be taken outside the horizon boundary integral. Denoting the horizon contribution to $B_{\partial\mathcal B}$ by $B_H$, the exterior normal $s^A=-r^A$ gives
\begin{equation}
	B_H=8G(\Omega_H-\Omega_\infty)J_H .
\label{eq:BHflux}
\end{equation}
The factor $8G$ reflects the $2\pi$ azimuthal integration in $J_H$, which is absent from the meridional balance. Appendix~\ref{app:horizon} verifies the required flux bound for the near-horizon expansions used below.

For vacuum Kerr in the nonrotating frame at infinity, the outer flux vanishes and the exact balance reduces to
\begin{equation}
	\int_\B W|\D\Omega|^2\dd V_q=8G\Omega_HJ_H.
\label{eq:Kerrbalance}
\end{equation}
The horizon supplies the entire positive balance and can therefore sustain frame dragging even in a vacuum exterior.

\subsection{Horizon regularity constrains scalar motion}

The Killing field $\ell^\mu=k^\mu+\Omega_Hm^\mu$ is both normal to the horizon and tangent to its null generators. Killing symmetry preserves the area and shape of the horizon cross sections, so the generators have zero expansion and shear. Their normality to the horizon also makes them twist-free. Raychaudhuri's equation therefore reads \cite{Raychaudhuri1955}
\begin{equation*}
	\ell^\mu\nabla_\mu\theta
	=\kappa\theta-\frac12\theta^2
	-\sigma_{\mu\nu}\sigma^{\mu\nu}
	-R_{\mu\nu}\ell^\mu\ell^\nu,
\end{equation*}
where $\ell^\nu\nabla_\nu\ell^\mu=\kappa\ell^\mu$ defines the surface gravity $\kappa$. Setting $\theta=0$ and $\sigma_{\mu\nu}=0$ therefore gives $R_{\mu\nu}\ell^\mu\ell^\nu=0$ on the horizon.

Contracting Einstein's equations with the null generator removes the metric-proportional trace term and gives
\begin{equation}
	T_{\mu\nu}\ell^\mu\ell^\nu=0 .
\label{eq:nullstress}
\end{equation}
For minimally coupled canonical real scalars, the potential also drops out of this null projection:
\begin{equation}
	T_{\mu\nu}\ell^\mu\ell^\nu
        =|\ell^\mu\partial_\mu\bm\varphi|^2 .
\label{eq:scalarhorizon}
\end{equation}
The right side is a sum of squares, so each scalar component must be constant along the horizon generators.

For fields time independent in the asymptotic frame, Eq.~\eqref{eq:inherit} reduces this horizon condition to
\begin{equation}
	0=(\Omega_H-\Omega_\infty)^2 |\partial_\phi\bm\varphi_H|^2 .
\label{eq:realhorizonlock}
\end{equation}
A nonzero angular derivative at any horizon point therefore fixes $\Omega_H=\Omega_\infty$. With uniform vanishing boundary values and no additional physical boundary, the finite-region criterion then excludes exterior frame dragging. If the angular derivative vanishes everywhere on the horizon, this identity gives no restriction on $\Omega_H$.

Table~\ref{tab:rotationmechanisms} summarizes the contrasting source and boundary mechanisms in these examples.

\begin{table*}[thb]
\caption{Source and boundary mechanisms for frame dragging. See Sec.~\ref{sec:finite} for geometric and boundary assumptions, Secs.~\ref{sec:real}--\ref{sec:BV} for scalar sources, and Sec.~\ref{sec:BH} for horizon conditions.}
\label{tab:rotationmechanisms}
\small
\begin{tabular}{@{}>{\raggedright\arraybackslash}p{0.145\textwidth}@{\hspace{0.035\textwidth}}
 >{\raggedright\arraybackslash}p{0.205\textwidth}@{\hspace{0.035\textwidth}}
 >{\raggedright\arraybackslash}p{0.19\textwidth}@{\hspace{0.035\textwidth}}
 >{\raggedright\arraybackslash}p{0.33\textwidth}@{}}
\hline\hline
Example & Momentum source & Boundary role & Consequence \\
\hline
Time-independent real scalars &
$j_\phi=-\Omhat S/N$, $S\geq0$ &
$\Omhat$ vanishes uniformly at all physical boundaries &
Frame dragging is excluded, irrespective of the scalar potential. \\
\hline
Harmonic complex scalar, $n\ne0$ &
The phase frequency can supply angular momentum &
Infinity gives $0$ and a synchronized horizon gives $\wh/n$ &
Rotation is compatible with the balance. Frame dragging is bounded by the phase-pattern and asymptotic values. \\
\hline
Helical global monopole, $n\ne0$ &
Conical infinity fixes $\omega=n\Omega_\infty$, giving $S=n^2F^2$ &
A synchronized horizon has $\Omhat_H=0$ &
Frame dragging is excluded throughout the connected horizonless region or synchronized exterior. \\
\hline
Vacuum Kerr &
$j_\phi=0$ in the exterior &
A spinning horizon supplies $B_H>0$ &
The boundary supports the positive rotational balance. \\
\hline\hline
\end{tabular}
\end{table*}

\subsection{Localized scalar hair can rotate}

For the harmonic complex scalar, the horizon condition Eq.~\eqref{eq:nullstress} becomes
\begin{equation}
	0=2(\omega-n\Omega_H)^2f_H^2 .
\label{eq:complexnull}
\end{equation}
If $f_H$ is nonzero at any point, then
\begin{equation}
	\omega=n\Omega_H,\qquad
	\wh=n(\Omega_H-\Omega_\infty).
\label{eq:complexsync}
\end{equation}
For $n\ne0$ the phase pattern co-rotates with the horizon. The current flux along it vanishes,
\begin{equation*}
	\ell_\mu\mathcal J^\mu=2(n\Omega_H-\omega)f_H^2=0
\end{equation*}
This is horizon synchronization \cite{Herdeiro2014,Smolic2017}.\footnote{For $n=0$ and $f_H\ne0$, the horizon identity fixes $\omega=0$ but does not constrain $\Omega_H$. If $f_H$ vanishes everywhere on the horizon, the identity gives no synchronization condition, and the scalar equation must be examined separately.}

For $n\ne0$, synchronization fixes the field's phase-pattern rate to the horizon angular velocity, while the localization condition \eqref{eq:frequencybound} permits both to rotate relative to infinity. Rotating black holes with synchronized complex-scalar hair realize this possibility \cite{Herdeiro2014}.

For a synchronized field, the angular-momentum density is
\begin{equation}
	j_\phi=\frac{2n^2}{N}(\Omega_H-\Omega)f^2 .
\label{eq:complexBHj}
\end{equation}
For $n\ne0$, when the inner and outer boundary values tend uniformly to $\Omhat_H=\wh/n$ and zero, the boundary-value argument of Sec.~\ref{sec:complex} gives Eq.~\eqref{eq:patternbound}. Frame dragging is bounded between its synchronized horizon and asymptotic values.

Let $\B_\varepsilon$ denote the meridional exterior truncated at a curve $\mathcal C_\varepsilon$ just outside the horizon, and let $B_{H,\varepsilon}$ be its inner-boundary contribution from Eq.~\eqref{eq:flux}. With the outer boundary flux removed, the balance is
\begin{equation}
	\begin{aligned}
		\int_{\B_\varepsilon}W|\D\Omhat|^2\dd V_q
		&=\int_{\B_\varepsilon}
		C\Omhat(\Omega_H-\Omega)\dd V_q
		+B_{H,\varepsilon},
	\end{aligned}
\label{eq:grouped}
\end{equation}
with $C$ from Eq.~\eqref{eq:complexdriven}. Under the preceding bound, $\Omhat(\Omega_H-\Omega) =\Omhat(\Omhat_H-\Omhat)\geq0$, so the scalar contribution is nonnegative. The horizon contribution also remains in the balance, and these terms can support nonzero frame dragging. The matter product must be kept intact because its separately expanded terms need not have convergent integrals. Equation~\eqref{eq:grouped} is an identity at finite horizon cutoff. Its full horizon limit requires convergence of the grouped matter integral. Appendix~\ref{app:horizon} establishes local convergence for the nonextremal expansions and explains why the extremal metric estimates alone do not establish it.

The same momentum constraint determines how the total angular momentum is divided between the field and the horizon. Integrating Eq.~\eqref{eq:Jdiv} over an asymptotically flat exterior gives the outer contribution $J$ and the inner contribution $-J_H$. With finite surface integrals and exterior phase charge, Eq.~\eqref{eq:Js} yields
\begin{equation}
	J-J_H=n\mathcal Q_{\rm ext}.
\label{eq:Jhole}
\end{equation}
Here $J$ is the total ADM angular momentum measured at spatial infinity, $J_H$ is the horizon angular momentum, and $\mathcal Q_{\rm ext}$ is the exterior phase charge. For a regular horizonless configuration, this inner boundary is absent, recovering $J=n\mathcal Q$.

\subsection{The monopole example with a horizon}

For the global helical field with winding $n\ne0$, Eq.~\eqref{eq:nullstress} gives
\begin{equation}
	0=(\omega-n\Omega_H)^2F_H^2 .
\label{eq:BVnull}
\end{equation}
If $F_H$ is nonzero at any point, then $\omega=n\Omega_H$. The conical boundary fixes $\omega=n\Omega_\infty$ through Eq.~\eqref{eq:syncinfty}, so
\begin{equation}
	\Omega_H=\frac{\omega}{n}=\Omega_\infty,\qquad \Omhat_H=0 .
\label{eq:bothsync}
\end{equation}
The persistent monopole field thus synchronizes the horizon with the asymptotic frame for every nonzero winding. For this synchronized exterior, $\Omhat$ vanishes at the horizon and tends to zero at infinity. With uniform boundary limits, the finite-region argument applied to Eq.~\eqref{eq:BVpositivePDE} yields $\Omhat=0$ throughout the exterior.

The distinction from localized scalar hair lies in the asymptotic boundary condition. Localized harmonic scalar hair can synchronize with a spinning horizon while rotating relative to infinity. The persistent monopole field locks its phase pattern to the asymptotic frame, and horizon synchronization then fixes the horizon to that frame as well.

Appendix~\ref{app:horizon} provides the boundary estimates and examines fields whose transverse amplitude vanishes everywhere on the horizon. It also gives the corresponding integral balance when the bulk integrals converge and the boundary flux vanishes.

\section{Conclusions}
\label{sec:conclusions}

We have derived an exact no-rotation criterion from the Einstein momentum constraint in circular, stationary, axisymmetric spacetimes. Frame dragging is excluded when the weighted matter contribution is nonpositive, the boundary flux vanishes, and the bulk integrals converge. For $\Omhat j_\phi\leq0$ pointwise, uniformly vanishing boundary values suffice, without infinite-volume integrals.

Time-independent, minimally coupled canonical real fields oppose frame dragging. A harmonic complex field can instead supply angular momentum, with its relative phase-pattern rate bounding horizonless frame dragging. For global helical BV fields, a persistent vacuum profile at a conical end locks the phase pattern to the asymptotic frame and excludes horizonless rotation for general meridional amplitudes and every nonzero winding. A nonzero transverse horizon amplitude also synchronizes the horizon with that frame and excludes exterior frame dragging. Future work includes other matter sources, self-gravitating multimonopole solutions, and extensions beyond circularity.

\begin{acknowledgments}
Qing-hai Wang thanks Prof.~Jiangbin Gong for useful discussions and helpful comments on the manuscript's presentation.

The AI tools identified in Sec.~\ref{sec:balance} also assisted with manuscript organization, language revision, clarification of physical arguments, and literature checks. The authors directed and reviewed all revisions, independently verified the scientific claims, checked references against the original sources, and take full responsibility for the manuscript.
\end{acknowledgments}

\section*{Data availability}
No data were created or analyzed in this study. The results follow from the analytical equations and arguments presented in the paper.

\appendix

\section{Boundary estimates and finite-region bounds}
\label{app:boundaries}

\subsection{Axis, center, and infinity}

The balance is first evaluated with finite outer and inner cutoffs. Near a regular axis, proper axis distance $\varrho$ gives $\Phi=\varrho+O(\varrho^3)$ and $\partial_\varrho\Omega=O(\varrho)$. Away from a horizon, $N$ has a finite positive limit, so the flux density $\Omhat W s^A\D_A\Omhat$ is $O(\varrho^4)$ for bounded $\Omhat$. At a regular center, $W=O(r^3)$ and the boundary line element is $O(r)$, giving an integrated flux $O(r^4)$ for bounded $\Omhat$ and gradient. Both contributions vanish. Here $\varrho$ and $r$ measure distance from the axis and center, respectively. Horizon limits are treated in Appendix~\ref{app:horizon}.

Smooth scalar modes with axial winding $n$ have $f,F=O(\varrho^{|n|})$ in regular Cartesian coordinates. Thus $n^2F^2/\Phi^2=O(\varrho^{2|n|-2})$ for $n\ne0$, and the axial-gradient energy remains regular.

At a flat or conical end, a sufficient falloff for the infinite-volume rotational balance is
\begin{equation}
	\begin{gathered}
		\Omhat=O(r^{-p}),\qquad \partial_r\Omhat=O(r^{-p-1}),\\
		\partial_\theta\Omhat=O(r^{-p}),\qquad p>3/2,
	\end{gathered}
\label{eq:outerfalloff}
\end{equation}
uniformly in angle, including regular axis limits. For $N\to N_\infty>0$, $\Phi=O(r)$, $q_{rr}=O(1)$, and $q_{\theta\theta}=O(r^2)$ with nondegenerate normalized limits, the outer flux is $O(r^{3-2p})$ and the gradient integral has radial weight $O(r^{2-2p})\dd r$. A bounded $S$ in Eq.~\eqref{eq:signsource} gives the same convergence bound for the scalar term. Other sources require their own estimate. The finite-region proof below instead uses boundary values and requires no specified decay rate.

\subsection{Finite-region bounds on frame dragging}
\label{app:comparison}

Use the geometry and boundary assumptions of Sec.~\ref{sec:finite}. On a connected finite meridional region $\B_R$, let $\delta$ bound $|z|$ on all imposed outer and inner boundaries and set $w\defeq(z-\delta)_+$. Multiplying Eq.~\eqref{eq:rotation} by $w$ and integrating by parts gives
\begin{equation}
	\int_{\B_R}W|\D w|^2\dd V_q
	=16\pi G\int_{\B_R}\Phi j_\phi w\dd V_q\leq0.
\label{eq:signcomparisonproof}
\end{equation}
The imposed boundary terms vanish because $w=0$ there. Axis and center cutoffs are removed at fixed outer and horizon cutoffs using the preceding estimates. Where $w>0$, $z>\delta\geq0$, so $zj_\phi\leq0$ gives $j_\phi\leq0$. Positivity, connectedness, and zero boundary values imply $w=0$, or $z\leq\delta$. Applying the argument to $-z$ and $-j_\phi$ gives $z\geq-\delta$, proving Eq.~\eqref{eq:bound}. Uniformly vanishing boundary values send $\delta$ to zero without an infinite-volume integral. No linear source law is needed. For Eq.~\eqref{eq:signsource}, the same calculation reads
\begin{equation}
	0=\int_{\B_R}\left(W|\D w|^2+czw\right)\dd V_q.
\label{eq:comparisonproof}
\end{equation}

For the harmonic field, write $p\defeq\widehat\Omega_{\rm p}$. If $p>0$, multiplication by $w_+\defeq(\Omhat-p-\delta)_+$ excludes $\Omhat>p+\delta$ in Eq.~\eqref{eq:patternPDE}, while multiplication by $-w_-$, with $w_-\defeq(-\Omhat-\delta)_+$, excludes $\Omhat<-\delta$. The nonnegative terms are respectively $W|\D w_+|^2+C(\Omhat-p)w_+$ and $W|\D w_-|^2+C(p-\Omhat)w_-$. Sign reversal covers $p<0$, and $p=0$ reduces to the opposing-source case. For a horizonless region choose $\delta=\sup_{\rm outer}|\Omhat|$. For a synchronized horizon use
\begin{equation}
	\delta_{R,\varepsilon} \defeq \max\left(\sup_{\rm outer}|\Omhat|,
    \sup_{\rm inner}|\Omhat-p|\right).
\label{eq:patterncutofferror}
\end{equation}
The resulting bounds are $\min(0,p)-\delta\leq\Omhat \leq\max(0,p)+\delta$. Uniform boundary limits give Eq.~\eqref{eq:patternbound}.

\section{Asymptotic decay of a massive harmonic scalar}
\label{app:decay}

For $U(s)=\mu^2s+O(s^2)$ and $f\to0$ in an asymptotically flat geometry, expand $f(r,\theta)\e^{\ri n\phi}=\sum_{l\geq|n|}R_l(r)Y_{ln}(\theta,\phi)$. Here $R_l$ is the radial amplitude and $\Delta_{S^2}Y_{ln}=-l(l+1)Y_{ln}$. The time and azimuthal derivatives of the harmonic phase combine into $(\omega-n\Omega)^2/N^2-n^2/\Phi^2$. Neglecting gravitational corrections, the asymptotic scalar equation therefore gives
\begin{equation}
	\begin{gathered}
		R_l'' + \frac2rR_l' - \left[\kappa_\infty^2 + \frac{l(l+1)}{r^2}\right] R_l = 0, 
	\end{gathered}
\label{eq:decayeq}
\end{equation}
where $\kappa_\infty^2\defeq\mu^2 - {\wh^2}/{N_\infty^2}$ and primes denote $r$ derivatives. Exponential decay requires $\kappa_\infty^2>0$, giving Eq.~\eqref{eq:frequencybound}. The leading flat-space tail is $\e^{-\kappa_\infty r}/r$. Gravitational corrections can change the multiplying power, and the threshold $\kappa_\infty=0$ requires separate analysis.

\section{From stationary BV stress to internal rotations}
\label{app:symmetry}

We derive the local helical field from stress symmetry, then state the matching requirements for a common global rotation. This conditional classification is independent of the application in Sec.~\ref{sec:BV}, which prescribes a global helical field. For the distinction between stress and field symmetries, see Ref.~\cite{Smolic2015}.

\subsection{Internal rotation where the field gradients are independent}

Define $Q_{\mu\nu}\defeq\partial_\mu\chi^a\partial_\nu\chi^a$,
$Q\defeq g^{\mu\nu}Q_{\mu\nu}$, and $X^a\defeq\pounds_{\bm\xi}\chi^a$ for a Killing field $\bm\xi$ preserving $T_{\mu\nu}$. Then
\begin{equation}
	\begin{aligned}
		\pounds_{\bm\xi}Q_{\mu\nu}
		&=\partial_\mu X^a\partial_\nu\chi^a
		+\partial_\mu\chi^a\partial_\nu X^a\\
		&=g_{\mu\nu}\left(\tfrac12\pounds_{\bm\xi}Q + \pounds_{\bm\xi}V\right).
	\end{aligned}
\label{eq:LieQ}
\end{equation}
On a connected region with three independent field gradients, let $v^\mu\partial_\mu\chi^a=0$ span their common constant-field direction and require $g(v,v)\ne0$. Contraction with $v^\mu v^\nu$ sets the parentheses to zero. Taking the trace gives
\begin{equation}
	\pounds_{\bm\xi}Q_{\mu\nu}=0,\qquad
	\pounds_{\bm\xi}Q=0,\qquad \pounds_{\bm\xi}V=0.
\label{eq:inheritQ}
\end{equation}
A contraction of the first line of Eq.~\eqref{eq:LieQ} with one $v^\mu$ also gives $v^\mu\partial_\mu X^a=0$. Thus locally $X^a=X^a(\bm\chi)$, and independence of the gradients reduces Eq.~\eqref{eq:inheritQ} to
\begin{equation}
	\partial_aX_b+\partial_bX_a=0,
	\qquad \partial_a\defeq\partial/\partial\chi^a.
\label{eq:internalKilling}
\end{equation}
Internal indices use $\delta_{ab}$. Differentiating and permuting the indices gives $\partial_a\partial_bX_c=0$, hence
\begin{equation}
	\bm X=\bm a+A\bm\chi,\qquad A^T=-A,
\label{eq:affine}
\end{equation}
with constant coefficients. Preservation of the BV potential requires $(|\bm\chi|^2-\eta^2)\bm a\cdot\bm\chi=0$. For $\bm a\ne0$, this confines the field image to a sphere or a plane, incompatible with three independent gradients. Therefore $\bm a=0$. Overlapping patches identify the same rotation generator throughout each connected region. The non-null requirement on $v$ is essential to the first contraction and is not implied by gradient independence.

\subsection{Commuting rotations and the helical field}

Applying the result to $\bm k$ and $\bm m$ gives
\begin{equation}
	\pounds_{\bm k}\bm\chi=A\bm\chi,\qquad
	\pounds_{\bm m}\bm\chi=B\bm\chi.
\label{eq:generators}
\end{equation}
Commuting spacetime derivatives give $(AB-BA)\bm\chi=0$. The three-dimensional field image implies $AB=BA$. In three internal dimensions, commuting rotation generators share an axis when $B\ne0$. Aligning it with the third direction gives $\partial_t\Xi=-\ri\omega\Xi$, $\partial_\phi\Xi=\ri n\Xi$, and $A=-(\omega/n)B$, including $A=0$. Integration yields
\begin{equation}
	\Xi=F(x^A)\e^{\ri[n\phi-\omega t+\alpha(x^A)]},
	\qquad \chi^3=H(x^A).
\label{eq:localhelix}
\end{equation}
Single-valuedness on a full axial orbit requires integer $n$ for a nonzero transverse field. The actual angular boundary profile fixes this integer, rather than its charge alone. Circularity then gives
\begin{equation}
	0=G_{A\phi}=8\pi G T_{A\phi}
	=8\pi G nF^2\partial_A\alpha.
\label{eq:phasecircular}
\end{equation}
For $n\ne0$, a constant internal rotation removes $\alpha$ on each connected patch where $F\ne0$.

\subsection{Matching the internal rotation between regions}

For a global classification, assume that the independent-gradient regions with non-null constant-field direction form a dense set. Each must be linked to the asymptotic region by a finite chain of shared interfaces with continuous fields and first derivatives. Across each interface, Eq.~\eqref{eq:generators} gives
\begin{equation}
	(A_+-A_-)\bm\chi=0,\qquad (B_+-B_-)\bm\chi=0.
\label{eq:matching}
\end{equation}
A nonzero antisymmetric matrix in three dimensions annihilates only one internal axis. Two noncollinear field values on each interface therefore make both generator differences zero. Successive matching fixes the asymptotic generators in every linked region, and continuity extends their relations to the remaining points, including boundaries where the common direction becomes null. This also excludes a linked region with $B=0$ when the asymptotic generator is nonzero.

Monopole charge alone does not guarantee these properties: zero interfaces, collinear field values, or open regions with dependent gradients can leave the matching undetermined. A single global real $F$ further requires compatible phase and sign choices between patches. The source and frequency arguments need only common generators and $F^2=|\Xi|^2$. Prescribing the global helical form supplies these generators without the classification assumptions.

\section{Monopole asymptotics and the internal frequency}
\label{app:monopole}

\subsection{Smooth angular profiles and monopole charge}
\label{app:windingprofiles}

On a reference time slice, write the unit vacuum field with polar profile $\Theta=\Theta(\theta)$ as
\begin{equation}
	\begin{aligned}
		\bm\sigma_\infty&\defeq\bm\chi_\infty/\eta\\
		&=\bigl(\sin\Theta\cos n\phi,\sin\Theta\sin n\phi,\cos\Theta\bigr).
	\end{aligned}
\label{eq:windingvacuum}
\end{equation}
Its topological charge is
\begin{equation}
	\begin{aligned}
		Q_{\rm top}&\defeq\frac1{4\pi}\int_{S^2}
		\bm\sigma_\infty\cdot (\partial_\theta\bm\sigma_\infty
        \times\partial_\phi\bm\sigma_\infty)\dd\theta\dd\phi\\
		&=\frac n2[\cos\Theta(0)-\cos\Theta(\pi)].
	\end{aligned}
\label{eq:monopolecharge}
\end{equation}
Thus opposite polar values $\Theta(0)=0$, $\Theta(\pi)=\pi$ give $Q_{\rm top}=n$, whereas equal polar values give zero. This charge is distinct from $\mathcal Q$ and is invariant under a constant proper internal rotation.

For $|n|>1$, the choice $\Theta=\theta$ is not smooth at the poles. A smooth family is
\begin{equation}
	\tan\frac{\Theta_n(\theta)}2
    =\left(\tan\frac\theta2\right)^{|n|}.
\label{eq:smoothwindingprofile}
\end{equation}
Near either pole, $\sin\Theta_n$ vanishes as the $|n|$th power of angular distance. The transverse field is a smooth multiple of $(X+\ri Y)^{|n|}$ for positive $n$, or its conjugate for negative $n$, in local Cartesian angular coordinates. For $n=1$ the standard hedgehog is recovered. On a unit round sphere the angular-gradient factor is
\begin{equation}
	\eta^2\left[(\Theta_n')^2
    +\frac{n^2\sin^2\Theta_n}{\sin^2\theta}\right].
\label{eq:windingangularenergy}
\end{equation}
It is angle dependent for $|n|>1$, while the unit hedgehog gives $2\eta^2$. These are smooth boundary maps, not constructed self-gravitating solutions. A compatible asymptotic geometry must satisfy the Einstein equations, and the standard unit-winding BV deficit cannot simply be carried over.

\subsection{Energy density at a conical end}

In a chart with $q_{r\theta}=0$, take
\begin{equation}
	\begin{gathered}
		q_{rr}=a_r(\theta)+h_r,\qquad
		q_{\theta\theta}/r^2=a_\theta(\theta)+h_\theta,\\
		\Phi/r=a_\phi(\theta)+h_\phi,
	\end{gathered}
\label{eq:conicalcoeffs}
\end{equation}
with smooth positive leading coefficients away from the regular axis. For some $\epsilon>0$, impose
\begin{equation}
	\partial_r^j\partial_\theta^k h_\alpha
	=O(r^{-\epsilon-j}),\qquad j+k\leq2,
\label{eq:metricfalloff}
\end{equation}
and the same bounds for $N-N_\infty$ and $\Omhat$. The bounds are uniform on angular patches, with their regular Cartesian versions at the axis. The additional spacetime condition $\Phi\Omhat\to0$ makes $k_\infty$ asymptotically timelike, but is not needed for the following energy estimate.

The spatial curvature is ${}^{(3)}R=O(r^{-2})$. Equation~\eqref{eq:Kphi} gives $K_{ij}K^{ij}=\Phi^2|\D\Omega|^2/(2N^2)=O(r^{-2\epsilon})$. The Hamiltonian constraint therefore yields
\begin{equation}
	16\pi G\rho={}^{(3)}R-K_{ij}K^{ij}\longrightarrow0.
\label{eq:energyconstraint}
\end{equation}
This establishes the geometric input used with the nonnegative BV energy in Sec.~\ref{sec:BV} to fix $\omega=n\Omega_\infty$. It allows angle-dependent leading coefficients and requires no scalar-derivative falloff estimates.

\section{Horizon expansions and exceptional amplitudes}
\label{app:horizon}

\subsection{Near-horizon geometry and rotational flux}

We use the following expansions in the circular exterior chart. For a nonextremal horizon, proper distance $\varrho_H$ and an angular coordinate $y$ give $q_{\varrho_H\varrho_H}=1$, $q_{\varrho_H y}=0$, and
\begin{equation}
	\begin{aligned}
		N&=\kappa_H\varrho_H+O(\varrho_H^3),&
		\Omega&=\Omega_H+O(\varrho_H^2),\\
		q_{yy}&=q_H(y)+O(\varrho_H^2),&
		\Phi&=\Phi_H(y)+O(\varrho_H^2).
	\end{aligned}
\label{eq:nonext}
\end{equation}
Here $\kappa_H>0$ is the surface gravity. For an extremal horizon use $x=r-r_H>0$, $q_{xy}=0$, and
\begin{equation}
	\begin{aligned}
		N&=n_H(y)x+O(x^2),&\Omega&=\Omega_H+O(x),\\
		q_{xx}&=Q_H(y)x^{-2}+O(x^{-1}),\\
		q_{yy}&=q_H(y)+O(x),&\Phi&=\Phi_H(y)+O(x).
	\end{aligned}
\label{eq:ext}
\end{equation}
The leading coefficients are smooth and positive away from the regular axis. Remainders are controlled through two radial and angular derivatives, with uniform regular-coordinate estimates at the axis endpoints. Scalar amplitudes are bounded. Unlike $\varrho_H$, $x$ is not proper distance: $q_{xx}\sim Q_H/x^2$ places the extremal horizon at infinite spatial distance. These exterior-chart expansions must be checked independently of smoothness in horizon-crossing coordinates.

Let $\mathcal C_\varepsilon$ approach the horizon and let $s^A$ point outward from the exterior, toward the hole. In the nonextremal case, $W=O(\varrho_H^{-1})$, $s^{\varrho_H}=-1$, and $\partial_{\varrho_H}\Omega=O(\varrho_H)$. In the extremal case, $W=O(x^{-1})$, $s^x=O(x)$, and $\partial_x\Omega=O(1)$. The flux density is bounded in both cases. The finite limiting line element and uniform axis estimates give
\begin{equation}
	\sup_{\varepsilon}\int_{\mathcal C_\varepsilon}
    |W s^A\D_A\Omega|\dd S_q<\infty.
\label{eq:absolutehorizonflux}
\end{equation}
Axis cutoffs are removed before the radial horizon limit. Replacing $\Omhat$ by $\Omhat_H$ in the weighted integral has error at most $\sup_{\mathcal C_\varepsilon} |\Omhat-\Omhat_H|$ times this absolute integral. Uniform convergence makes that error vanish. A finite signed flux limit defining $J_H$ remains a separate assumption.

\subsection{Synchronized monopole boundary values and flux}

Once $\Omega_H=\Omega_\infty$, Eq.~\eqref{eq:nonext} gives $\Omhat=O(\varrho_H^2)$ and $\Omhat W s^A\D_A\Omhat=O(\varrho_H^2)$. Equation~\eqref{eq:ext} instead gives $\Omhat=O(x)$ and a weighted flux density $O(x)$. Thus the inner boundary values vanish uniformly and the integrated horizon flux vanishes in both cases.

The gradient and scalar terms are locally integrable, with nonextremal radial weights $O(\varrho_H)\dd\varrho_H$ and $O(\varrho_H^3)\dd\varrho_H$, respectively. Both extremal weights are at worst $O(1)\dd x$. Regular-axis behavior controls the angular endpoints. If the outer flux vanishes and the integrals also converge at infinity, the balance is
\begin{equation}
	\int_\B\left[W|\D\Omhat|^2
	+16\pi G\frac{n^2\Phi F^2}{N}\Omhat^2\right]\dd V_q=0.
\label{eq:BVidentity}
\end{equation}
The finite-region proof needs only the boundary values.

\subsection{A transverse field vanishing everywhere on the horizon}

When $F_H\equiv0$, Eq.~\eqref{eq:BVnull} gives no frequency restriction. The transverse amplitude equation is
\begin{equation}
	\begin{aligned}
		&\frac{\partial_A(N\Phi\sqrt q\,q^{AB}\partial_BF)}{N\Phi\sqrt q}\\
		&\quad+\left[\frac{(\omega-n\Omega)^2}{N^2}
		-\frac{n^2}{\Phi^2}-\lambda(F^2+H^2-\eta^2)\right]F=0.
	\end{aligned}
\label{eq:BVamplitude}
\end{equation}
On an angular patch away from the axis assume a first nonzero power, with $\zeta=\varrho_H$ or $x$:
\begin{equation}
	\begin{gathered}
		F=\zeta^p f_p(y)+R,\qquad p>0,\quad f_p\not\equiv0,\\
		\partial_\zeta^j\partial_y^kR=o(\zeta^{p-j}),\qquad j+k\leq2,
	\end{gathered}
\label{eq:Ffinite}
\end{equation}
uniformly on the patch, with smooth $f_p$. Set $\Delta_H\defeq\omega-n\Omega_H$.

The nonextremal radial operator is $\partial_{\varrho_H}^2+\varrho_H^{-1}\partial_{\varrho_H}$. Its leading term combines with the frequency term to give
\begin{equation}
	\left(p^2+\frac{\Delta_H^2}{\kappa_H^2}\right)
	f_p\varrho_H^{p-2}+o(\varrho_H^{p-2})=0.
\label{eq:nonextF}
\end{equation}
Angular, axial, and potential terms are higher order. The positive coefficient excludes this finite-order branch even when $\Delta_H=0$.

At an extremal horizon the radial operator is $Q_H^{-1}(x^2\partial_x^2+2x\partial_x)$, so its action and the angular, axial, and potential terms are $O(x^p)$. For $\Delta_H\ne0$, the more singular frequency term gives
\begin{equation}
	\frac{\Delta_H^2}{n_H^2}f_p x^{p-2}+o(x^{p-2})=0,
\label{eq:extF}
\end{equation}
forcing $\Delta_H=0$. With synchronization, all leading terms are $O(x^p)$ and a vanishing branch is not excluded. Together with $\omega=n\Omega_\infty$, this yields $\Omega_H=\Omega_\infty$ and permits the finite-region no-rotation proof. Existence of such a branch remains to be determined. Fields vanishing faster than every power, or zero throughout a horizon neighborhood, are not covered.

\subsection{The grouped complex-scalar balance}

For a synchronized complex field, the matter product in Eq.~\eqref{eq:grouped} is $C\Omhat(\Omega_H-\Omega)$. At a nonextremal horizon, $C=O(\varrho_H^{-1})$ and $\Omega_H-\Omega=O(\varrho_H^2)$, giving an integrable $O(\varrho_H)$ product with finite meridional density. Splitting it as
\begin{equation}
	C\Omhat(\Omega_H-\Omega)
	=C\Omhat_H\Omhat-C\Omhat^2
\label{eq:split}
\end{equation}
can instead produce two logarithmically divergent integrals when $f_H\ne0$ and $\Omhat_H\ne0$. Their leading terms cancel at a common cutoff, so the product must be integrated before the horizon limit. At an extremal horizon, Eq.~\eqref{eq:ext} permits an $O(1)$ product against $\dd x/x$. These metric estimates alone do not establish convergence, which requires a separate check using the field and metric behavior.

\end{document}